# Giant gate-controlled proximity magnetoresistance in semiconductor-based ferromagnetic/nonmagnetic bilayers

Kosuke Takiguchi[1], Le Duc Anh[1,2,*], Takahiro Chiba[3], Tomohiro Koyama[4], Daichi Chiba[4], and Masaaki Tanaka[1,5,*]

[1] *Department of Electrical Engineering and Information Systems, The University of Tokyo, Bunkyo-ku, Tokyo 113-8656, Japan.*
[2] *Institute of Engineering Innovation, The University of Tokyo, Bunkyo-ku, Tokyo 113-8656, Japan.*
[3] *National Institute of Technology, Fukushima College, Iwaki, Fukushima, 970-8034, Japan*
[4] *Department of Applied Physics, The University of Tokyo, Bunkyo-ku, Tokyo 113-8656, Japan*
[5] *Center for Spintronics Research Network, The University of Tokyo, Bunkyo-ku, Tokyo 113-8656, Japan.*
\* Email: anh@cryst.t.u-tokyo.ac.jp
masaaki@ee.t.u-tokyo.ac.jp

The evolution of information technology has been driven by the discovery of new forms of large magnetoresistance (MR), such as giant magnetoresistance (GMR)[1,2] and tunnelling magnetoresistance (TMR)[3,4] in magnetic multilayers. Recently, new types of MR have been observed in much simpler bilayers consisting of ferromagnetic (FM)/nonmagnetic (NM) thin films[5-10]; however, the magnitude of MR in these materials is very small (0.01 ~ 1%). Here, we demonstrate that NM/FM bilayers consisting of a NM InAs quantum well conductive channel and an insulating FM (Ga,Fe)Sb layer exhibit giant proximity magnetoresistance (PMR) (~80% at 14 T). This PMR is two orders of magnitude larger than the MR observed in NM/FM bilayers reported to date, and its magnitude can be controlled by a gate voltage. These results are explained by the penetration of the InAs two-dimensional-electron **wavefunction into (Ga,Fe)Sb. The ability to strongly modulate the NM channel current by both electrical and magnetic gating represents a new concept of magnetic-gating spin transistors.**

High-quality semiconductor materials are at the core of high-performance electronic devices, such as transistors and lasers, due to their high carrier mobility and coherency. However, the volatility of information stored in semiconductor structures inevitably causes large power consumption in both operation and standby modes, which rises rapidly with the degree of device integration and miniaturization. Introducing non-volatility into semiconductors is thus strongly desired, and ferromagnetic semiconductors (FMS) have been successfully realized by doping a large amount (from one to tens of percent) of magnetic impurities into NM semiconductors, which solves the problem of volatility[11-14]. Such incorporation of magnetic impurities, however, has major unwanted impacts on the properties of the host semiconductors, significantly degrading the carrier mobility and coherency and even forming impurity bands that invalidate the conventional understanding of well-established semiconductor materials and devices[14]. Achieving a very large magnetic response in a NM semiconductor without magnetic impurity doping is therefore of crucial importance; however, doing so is highly challenging.

Utilizing the magnetic proximity effect (MPE) is a promising pathway to avoid the aforementioned dilemma in FMSs. The MPE, which occurs due to the magnetic coupling within a few monolayers at the interface of two magnetically dissimilar layers, is effective in incorporating magnetic effects into NM materials in which doping magnetic impurities is difficult or undesirable[15,16]. In particular, the MPE at FM/NM interfaces where the NM layer is a material with strong spin-orbit coupling—typically a heavy metal, TI, or a two-dimensional electron gas (2DEG) with a strong Rashba effect—has attracted much attention as these materials represent promising components for spintronics devices and as a route toward realizing Majorana fermions[17,18]. Using such FM/NM interfaces, several groups have reported new proximity-originated MR effects[5-10], which have led to a new ability to manipulate electron transport in the NM channel by the magnetization of the FM layer. However, the MR ratio is too small (0.01 ~ 1%) to consider any application, and the MR and related properties cannot be controlled by external means, such as a gate voltage, because of the short-range magnetic coupling (~1 nm) at these FM/NM interfaces and the short Thomas-Fermi screening length (~0.1 nm) in the metallic NM channels.

Here, we show that the magnetic coupling range of MPE in a NM/FM bilayer can be dramatically enhanced, at least by two orders of magnitude, to several tens of nm on the NM side by using a semiconductor quantum well (QW) as the NM layer. Due to the very high carrier coherency in single-crystal semiconductors (which can exceed 100 nm in a good sample), the electronic states in nm-scale semiconductor thin films are quantized, and all the electron carriers behave *collectively* in the quantized direction, which is perpendicular to the NM/FM interface. Therefore, even if the 2D electron wavefunction in the NM QW penetrates moderately (only a few nm) into the neighbouring FM layer, the magnetic coupling is effectively felt by *the whole* electronic system. Moreover, we can shift the position of the wavefunction by applying a gate voltage, through which the MPE can be largely controlled. As a proof-of-concept system, we studied bilayers consisting of NM InAs QW and FM (Ga,Fe)Sb (Fig. 1a). This bilayer system has several unique properties that are particularly suitable for demonstrating MPE: i) (Ga,Fe)Sb is a p-type FMS with a high Curie temperature $T_C$ (> 300 K)[19,20], while InAs QW is a typical Rashba 2DEG system with high electron mobility. ii) The lattice mismatch between InAs and (Ga,Fe)Sb is only ~0.1%[19]; thus, high-quality heterostructures can be epitaxially grown. iii) InAs/(Ga,Fe)Sb is a type-III heterostructure, i.e., the conduction band bottom of InAs is lower than the valence band top of (Ga,Fe)Sb at the NM/FM interface, which enables large penetration of the electron wavefunction into the (Ga,Fe)Sb side. iv) The resistivity of the (Ga,Fe)Sb layer is two orders of magnitude higher than that of the InAs 2DEG, particularly at low temperature [see Supplementary Information (S. I.)]. Therefore, the electrical transport in these bilayers is accounted for almost entirely (> 99%) by the InAs 2DEG. In this work, we show that giant proximity magnetoresistance (PMR) (~80% at 14 T) is induced in the InAs 2DEG due to the MPE from the neighbouring (Ga,Fe)Sb layer and that the PMR can be controlled over one order of magnitude using a gate voltage.

We grew heterostructures consisting of (from top to bottom) InAs (thickness $d_{InAs}$ nm)/(Ga,Fe)Sb (15 nm, Fe 20%, $T_C$ > 300 K)/AlSb (300 nm)/AlAs (15 nm)/GaAs (100 nm) on semi-insulating GaAs (001) substrates by molecular beam epitaxy (see Methods). The thickness $d_{InAs}$ was varied from 15 nm to 40 nm in a wedge-shaped sample to investigate the channel thickness dependence (see S. I.). We patterned 50 × 200 μm$^2$ Hall bars and measured magnetotransport properties by the standard 4-terminal method. First, we measured the MR of sample A with $d_{InAs}$ = 15 nm when a magnetic field **H** was applied parallel and perpendicular to the film plane (Fig. 1c). Our Hall measurements on this sample indicate n-type conduction and a mobility of 938 cm$^2$/Vs at 2 K for electron carriers in the InAs channel. We note that this mobility is impressively high considering the low growth temperature of the InAs layer (~235℃), which was required to maintain the crystal quality of the underlying FMS (Ga,Fe)Sb layer. The favourable quality of the InAs channel was further manifested by strong Shubnikov-de Haas (SdH) oscillations in the resistance when applying **H** perpendicular to the film plane (Fig. 1c), which remained clearly visible at temperatures of up to 100 K (Fig. 1d), indicating that the carrier coherency was preserved even at such high temperature. The SdH oscillations disappeared when applying an in-plane **H**, which is evidence of the formation of a 2DEG in the InAs QW.

In all the samples, we observed very large MR values of several tens of percent (for example, the MR value in sample A was 20% at 2 K and 10 T, as shown in Fig. 1c), which is unprecedented for NM InAs channels. To understand this unique MR, which is masked by SdH oscillations, we studied another sample (sample B, $d_{InAs}$ = 15 nm) with same structure as sample A but with the lower electron mobility of InAs (= 272 cm$^2$/Vs); thus, the sample did not show SdH oscillations (see Method). Figure 2a shows the angular dependence of the MR at 2 K in sample B when applying an external magnetic field **H** in various directions in the *x-y, y-z* and *z-x* planes, denoted as MR($\alpha$), MR($\beta$), and MR($\gamma$), respectively. The rotation angles $\alpha$, $\beta$ and $\gamma$ are defined as shown in the insets of Fig. 2a. The MR($\alpha$), MR($\beta$), and MR($\gamma$) plotted in Fig. 2 are given by $\Delta R/R_{min}$= [$R(\alpha, \beta, \gamma)$ − $R_{min}$]/$R_{min}$ at various magnetic field strengths (0 T ≤ $\mu_0 H$ ≤ 14 T) at 2 K, and the current $J_{DS}$ = 1 µA, where $R$ is the resistance, $\Delta R$ is the resistance change when the **H** rotation angles $\alpha$, $\beta$ and $\gamma$ are changed, and $R_{min}$ is the minimum resistance value. While MR($\alpha$) shows a small change ($\Delta R/R_{min}$ ~3%), the changes in MR($\beta$) and MR($\gamma$) reach ~80% at 14 T. These MR values observed in the InAs/(Ga,Fe)Sb bilayer are two to three orders of magnitude larger than those reported previously in other FM/NM bilayers[5-10]. In addition, the symmetry of the angular dependence [MR($\beta$) = MR($\gamma$) >> MR($\alpha$)] is completely different from that of the spin Hall MR (SMR) [MR($\alpha$) = MR($\beta$) >> MR($\gamma$)][5,6] and anisotropic MR (AMR) [MR($\alpha$) = MR($\gamma$) >> MR($\beta$)]. This finding indicates that the origin of the MR observed in InAs/(Ga,Fe)Sb is different from that of SMR and AMR. We note that although the current flows only in the NM InAs QW, the MR − H curves show clear hysteresis characteristics, as demonstrated in Fig.2b, in which the two peak positions agree with the coercivity of the (Ga,Fe)Sb layer. This finding indicates that the MPE at the interface of the InAs QW and the (Ga,Fe)Sb layer is the origin of this new MR, which we refer to as "proximity magnetoresistance (PMR)" hereafter. Furthermore, we observed this large PMR in several samples with both high and low electron mobilities (272 ~ 938 cm$^2$/Vs) up to a relatively high temperature (~100 K), meaning that quantum transport, such as weak localization in InAs[21], is *not* the origin of the PMR.

To quantify the MPE of the (Ga,Fe)Sb magnetization on the magnetotransport in the InAs QW, we reproduced the PMR − H characteristics using a phenomenological model, which is based on the model of Khosla and Fischer[22] but modified for the case of interfacial magnetic scattering. In a system with localized magnetic scattering centres, the MR can be described as follows:

$$\frac{\Delta \rho}{\rho} = -a^2 \ln(1 + b^2 H^2) + \frac{c^2 H^2}{1 + d^2 H^2}, \qquad (1)$$

where *a, b, c,* and *d* are fitting parameters defined in the Methods section. The first term, which gives a negative MR component (green dashed curves in Fig. 3a), is due to the Kondo scattering of electron carriers in the InAs QW by the localized spins at the (Ga,Fe)Sb interface. The second term, which gives a positive MR component (blue dashed curves in Fig. 3a), depends on the degree of *s-d* orbital hybridization between the electron carriers (*s* orbitals) in InAs and the localized spins (*d* orbitals) in (Ga,Fe)Sb at the bilayer interface due to the penetration of the 2DEG wavefunction into the FM side. This interfacial *s-d* hybridization induces the difference in electron density and mobility between the majority and minority spins at the Fermi level in InAs, whose effect is expressed in the parameter *d* of eq. (1):

$$d^2 = \frac{(\sigma_1 \mu_2 - \sigma_2 \mu_1)^2}{(\sigma_1 + \sigma_2)^2}, \qquad (2)$$

where $\sigma_i$ and $\mu_i$ are the conductivity and mobility of carriers, respectively, and $i = 1, 2$ denotes the majority and minority spins, respectively. Note that if there is no *s-d* orbital hybridization ($d = 0$), the second term of eq. (1) is simply the classic quadratic positive MR induced by the Lorentz force in NM conductors. The fitting of eq. (1) to our PMR data in Fig. 3a gives $d = 6.94$, which is significantly larger than the $d$ values obtained in magnetic-impurity-doped semiconductors ($d = 2.03 \times 10^{-5}$ in CdS[22] and 0.101 in InMnSb[23]). This finding implies a very strong interfacial *s-d* hybridization at the interface of our InAs/(Ga,Fe)Sb bilayers.

The other important advantage of the PMR in the InAs/(Ga,Fe)Sb bilayers over any other types of MR reported in NM/FM bilayers[5-10] is the controllability attained with a gate voltage $V_g$, by which we modulate the penetration of the electron wavefunction in the InAs QW into the underlying (Ga,Fe)Sb layer. As shown in Fig. 3b, the PMR ratio in a field-effect transistor (FET) fabricated on sample B (see Methods) can be varied by ten-fold at $\mu_0|H| < 1$ T when $V_g$ is varied from 3 V to −3 V. By fitting eq. (1) to the MR data (solid black curves in Fig. 3b), we obtained the value of parameter $d$ versus $V_g$ as shown in Fig. 3c. When $V_g = -3$ V, $d$ reaches a surprisingly large value of 58.5, which is $10^6$ times larger than that of CdS[22], indicating a strong enhancement of the *s-d* orbital hybridization by applying a negative $V_g$. We calculated the movement of the wavefunction in the InAs QW with $V_g$ by performing a self-consistent calculation of the electronic structure in the InAs/(Ga,Fe)Sb bilayer, as shown in Fig. 3d for $V_g = 3$ V and −3 V (see Methods). Using the calculated results, we quantify the degree of the wavefunction penetration in (Ga,Fe)Sb by

$$P = \int_{x_{(Ga,Fe)Sb}} |\varphi(x)|^2 dx, \qquad (3)$$

where $x_{(Ga,Fe)Sb}$ and $\varphi(x)$ denote the (Ga,Fe)Sb region and the electron wavefunction, respectively. Note that our calculations indicate that only the first subband of the QW is occupied. As expected, $P$ strongly increases with negative $V_g$. The $V_g$-dependences of $P$ (green triangles) and the PMR ratio (red circles) show a strong positive correlation (Fig. 3e), which confirms our aforementioned reasoning that the modulation of the PMR ratio is caused by the shift of the carrier wavefunction in the InAs QW with $V_g$.

Finally, we discuss the device application prospects of this new PMR. As shown in Fig. 4, by maintaining the high quality of the NM semiconductor channel, we have successfully realized FET operations in the InAs/(Ga,Fe)Sb bilayers; the current is modulated by over 2250% by varying $V_g$ from −3 V to +3 V. In addition, the current is altered by ~8% due to the PMR when applying $\mu_0 H = 1$ T to rotate the magnetization of (Ga,Fe)Sb from the in-plane to the perpendicular direction (the red and blue curves in Fig. 4a, b, respectively). Therefore, the device shows promise as a magnetic-gating spin FET if we can switch the magnetization direction of (Ga,Fe)Sb in a non-volatile manner. Furthermore, as InAs can also host superconductivity by the proximity effect with s-wave superconductors (SC)[24-27], the large and tuneable MPE in InAs/(Ga,Fe)Sb provides another important ingredient that makes SC/InAs/(Ga,Fe)Sb heterostructures a good candidate as a platform for realizing Majorana fermions[17,18].

## Methods

**Sample preparation and characterization**

We grew heterostructures consisting of InAs (thickness $d_{InAs}$ nm, 15 ~ 40 nm)/(Ga,Fe)Sb (15 nm, Fe 20%, $T_C$ > 300 K)/AlSb (300 nm)/AlAs (15 nm)/GaAs (100 nm) on semi-insulating GaAs (001) substrates by molecular beam epitaxy (MBE). The growth temperature ($T_S$) was 550℃ for the GaAs and AlAs layers, 470℃ for the AlSb layer, 250℃ for the (Ga,Fe)Sb layer, and 235℃ for the InAs layer. *In situ* reflection high energy electron diffraction (RHEED) patterns of InAs and (Ga,Fe)Sb are bright and streaky, indicating good crystal quality and smooth surface (see Supplementary Fig. S2b in S.I.). We intentionally controlled the mobility of the InAs layer by slightly increasing the As flux during the growth of this layer in sample B, which introduced more antisite defects into the crystal. A high-resolution scanning transmission electron microscopy (STEM) lattice image of the InAs/(Ga,Fe)Sb heterostructure reveals a high-quality single-phase zinc-blende-type crystal structure and a sharp interface between the NM InAs QW channel and the FM (Ga,Fe)Sb layer (Fig. 1b, leftmost panel). Energy dispersive X-ray spectroscopy (EDX) mapping of atomic distributions in the same heterostructure indicates that there is no Fe diffusion from the (Ga,Fe)Sb layer into the InAs layer (Fig. 1b, right panels).

**Fabrication process of the FET devices and transport measurement**

We first patterned the samples onto 50 x 200 μm$^2$ Hall bars by standard photolithography and Ar ion milling and then formed several electrodes (source S, drain D, and electrodes for transport measurements) via the electron-beam evaporation and lift-off of a Au (50 nm)/Cr (5 nm) film. An insulating HfO$_2$ layer (~50 nm) was deposited on the Hall bars by atomic layer deposition; finally, we formed a top gate electrode (G), again by electron beam evaporation and lift-off of a Au (50 nm)/Cr (5 nm) film. Fig. 1a (right panel) shows an optical microscope image of the FET device examined in this study. We applied the gate voltage $V_g$ between the G electrode and the D electrode. We measured the PMR by the standard 4-terminal method, using a Quantum Design physical property measurement system (PPMS) equipped with a rotating sample stage and a magnetotransport measurement system.

**Modified Khosla-Fischer model**

Equation (1) is based on the Khosla-Fischer model that describes magnetotransport phenomena in systems containing localized magnetic moments[22]. The parameters in eq. (1) are given by

$$a = A_1 J D(\epsilon_F)[S(S+1) + \langle M^2 \rangle], \qquad (4)$$

$$b^2 = \left[1 + 4S^2\pi^2 \left(\frac{2JD(\epsilon_F)}{g}\right)^4\right]\left(\frac{g\mu_B}{\alpha k_B T}\right), \qquad (5)$$

$$c^2 = \frac{\sigma_1 \sigma_2 (\mu_1 + \mu_2)^2}{(\sigma_1 + \sigma_2)^2}, \qquad (6)$$

$$d^2 = \frac{(\sigma_1 \mu_2 - \sigma_2 \mu_1)^2}{(\sigma_1 + \sigma_2)^2}. \qquad (7)$$

In eq. (4) and (5), $A_1$ is a constant representing the contribution of spin scattering to the whole MR, $\mu_B$ is the Bohr magneton, $\alpha$ is a numerical factor that is on the order of unity, $D(\epsilon_F)$ is the density of states (DOS) at the Fermi level, $g$ is the effective Lande factor of the InAs QW, $\langle M^2 \rangle$ is the averaged squared magnetization, $S$ is the localized spin moment of (Ga,Fe)Sb (we assume $S = 5/2$ for $Fe^{3+}$ ions in (Ga,Fe)Sb), and $J$ is the $s,p$-$d$ exchange interaction energy at the bilayer interface. In eq. (6) and (7), $\sigma_i$ and $\mu_i$ represent the conductivity and mobility of electron carriers in InAs, respectively. The subscripts 1 and 2 of each parameter denote the majority and minority spins, respectively.

**Self-consistent calculation of the electronic structure in InAs/(Ga,Fe)Sb bilayers**

We performed self-consistent calculations to obtain the $V_g$ dependence of the band profile and carrier distribution in InAs QW/(Ga,Fe)Sb bilayers using Schrödinger's equation (8) and Poisson's equation (9):

$$\left(\frac{-\hbar^2}{2m^*}\frac{\partial^2}{\partial z^2} + V_{\text{charge}}(z) + V_{\text{offset}}(z) + V_{\text{xc}}(z) + V_{\text{gate}}(z)\right)\varphi(z) = E\varphi(z) \qquad (8)$$

$$\frac{\partial^2}{\partial z^2}V_{\text{charge}} = e\frac{\rho_e(z)}{\epsilon} \qquad (9)$$

Here, $z$ is the growth direction, $V_{\text{charge}}$ is the space charge potential induced by electron carriers, $V_{\text{offset}}$ is the conduction band offset between InAs and GaSb at the Γ point (= 813 meV), $V_{\text{xc}}$ is the exchange-correlation potential of electrons[28], $V_{\text{gate}}$ is the potential induced by the gate voltage, and $\varphi(z)$ is the electron wavefunction. For InAs, the dielectric constant value $\varepsilon = 12.37\varepsilon_0$ and the electron effective mass at the Γ point $m^* = 0.08\ m_0$ were used. In these calculations, we assumed that the Fermi level is pinned at the valence band top of (Ga,Fe)Sb due to the formation of an impurity band there, and we used the carrier (electron) concentration measured by the Hall effect at every $V_g$ (see S. I.).

**Acknowledgments:** A part of this work was conducted at Advanced Characterization Nanotechnology Platform of the University of Tokyo, supported by "Nanotechnology Platform" of the Ministry of Education, Culture, Sports, Science and Technology (MEXT), Japan.

**Funding:** This work was partly supported by Grants-in-Aid for Scientific Research (Nos. 16H02095, 17H04922, 18H05345), the CREST Program (JPMJCR1777) of the Japan Science and Technology Agency, Yazaki Memorial Foundation for Science & Technology, and the Spintronics Research Network of Japan (Spin-RNJ).

**Author contributions:**

K. T. and L. D. A. designed the experiments and grew the samples. K. T. performed sample characterizations and transport properties. K. T., D. C., and T. K. fabricated the FET devices. K. T., L. D. A., and T. C. discussed on the mechanism and performed theoretical calculations. K. T., L. D. A. and M. T. planned the study and wrote the manuscript.

**Competing interests:** Authors declare no competing interests.

**Data and materials availability:** All data are available in the main text or the Supplementary Information.

**Figures and captions**

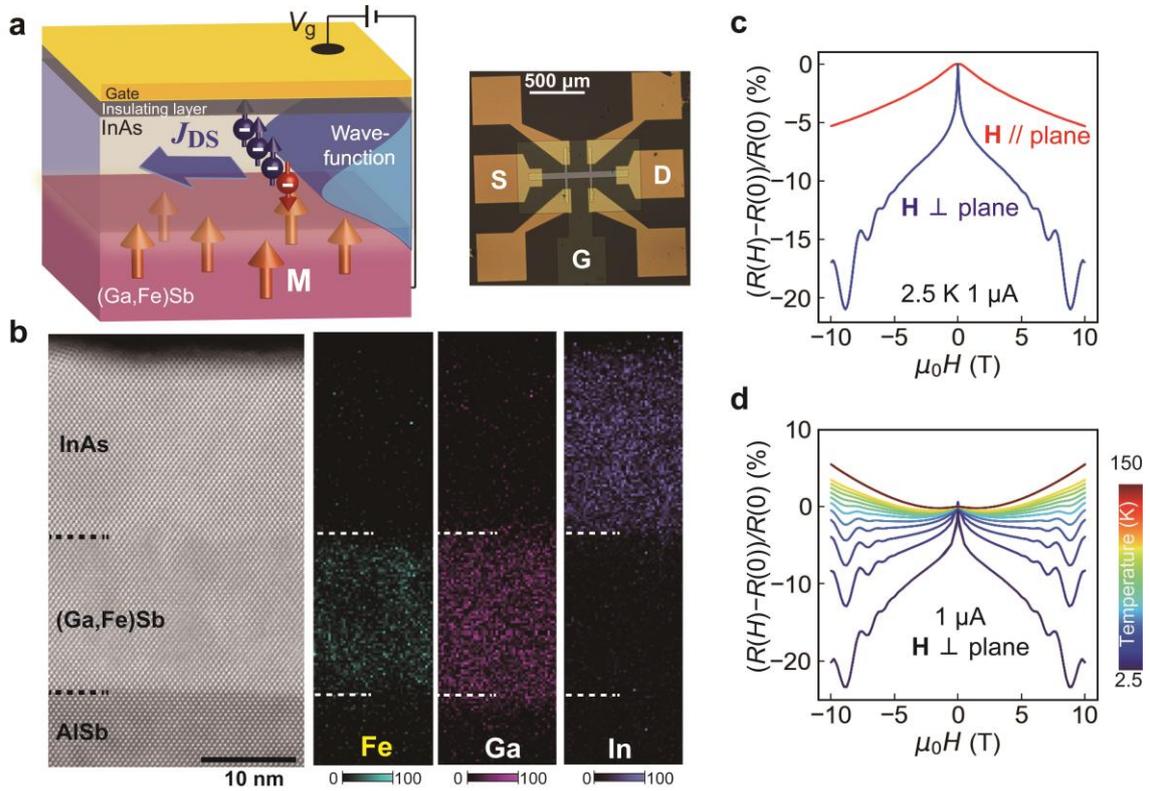

**Fig. 1| Device structure, microstructure characterization, and magnetotransport. a,** Schematic structure (left panel) and optical microscopy image (right panel) of the device examined in this study, which contains an NM InAs QW interfaced with an insulating FM (Ga,Fe)Sb. The left panel shows the results obtained by applying an electron current ($J_{DS}$) between the source (S) and the drain (D) electrode and the gate voltage $V_g$ between the gate (G) and the D electrode. Penetration of the electron wavefunction in InAs into (Ga,Fe)Sb (pink area in the left panel) induces a large proximity magnetoresistance (PMR) in the nonmagnetic InAs layer. By applying $V_g$ to control the electron wavefunction in InAs, we can significantly modulate the PMR. **b,** Scanning transmission electron microscopy (STEM, leftmost panel) and energy dispersive X-ray spectroscopy mapping of the Fe, Ga, and In distributions (right panels), respectively, in an InAs/(Ga,Fe)Sb with $d_{InAs}$ = 15 nm. **c,** Magnetoresistance with a magnetic field $\mu_0 H$ applied perpendicular (blue curve) and parallel (red curve) to the film plane. **d,** Temperature dependence of the magnetoresistance with $\mu_0 H$ applied perpendicular to the film plane.

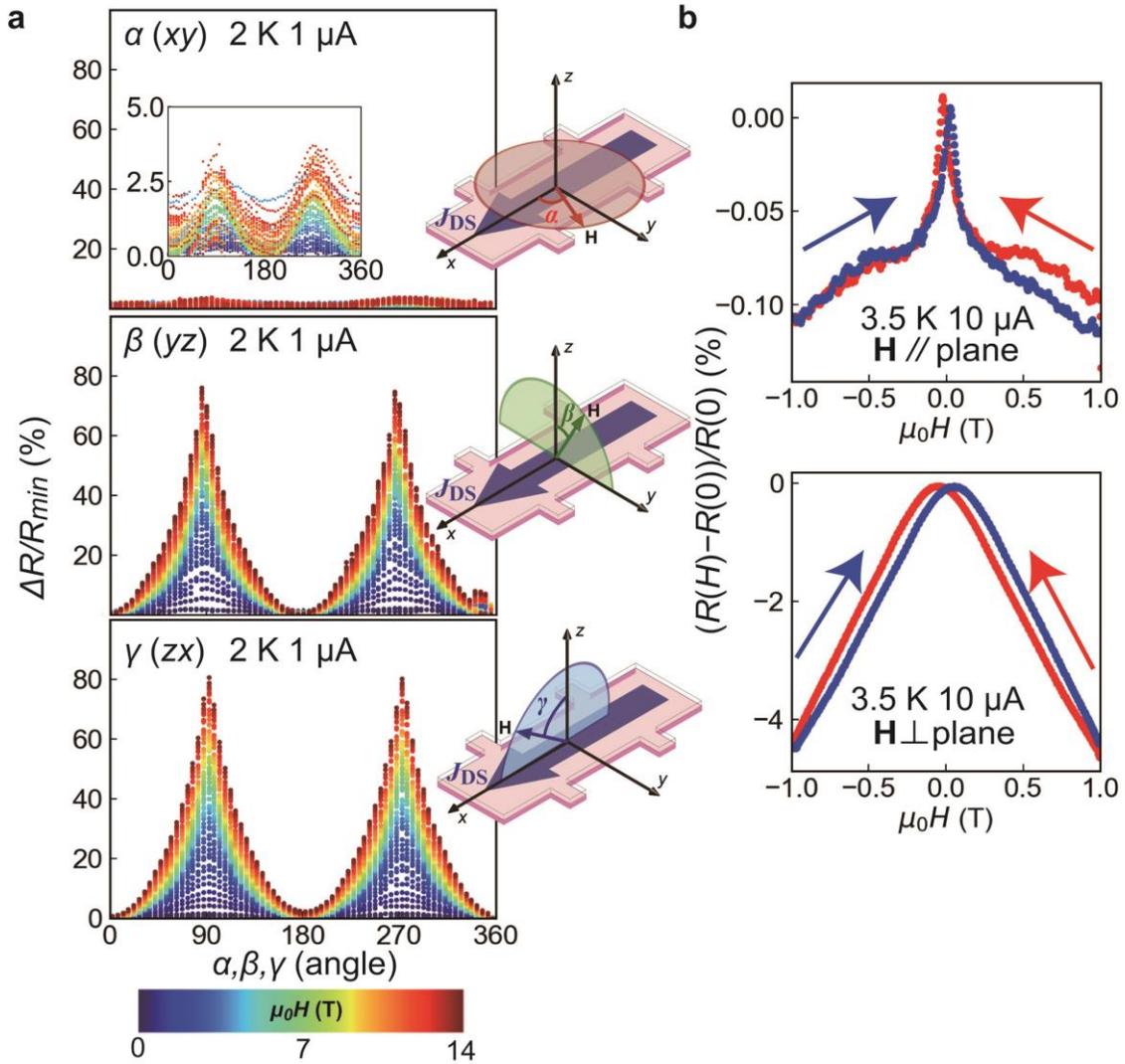

**Fig. 2| Dependence of PMR on the magnetic field direction and strength. a**, Magnetic field direction dependence of the PMR ratio, defined as $\Delta R/R_{min}= [R(\alpha, \beta, \gamma) - R_{min}]/ R_{min}$, under various magnetic field strengths (0 T ≤ $\mu_0 H$ ≤ 14 T) at 2 K and $J_{DS}$ = 1 μA. $R_{min}$ is the minimum value of the resistance under constant field rotation. The rotation angles ($\alpha, \beta, \gamma$) are defined as illustrated in the insets on the right side. The inset inside the top panel ($\alpha$ rotation) is a magnified plot of the PMR data. **b**, Magnetic field dependence of the magnetoresistance, plotted over the range of ±1 T, when the magnetic field is applied parallel (upper) and perpendicular (lower) to the film plane at 3.5 K and $J_{DS}$ = 10 μA.

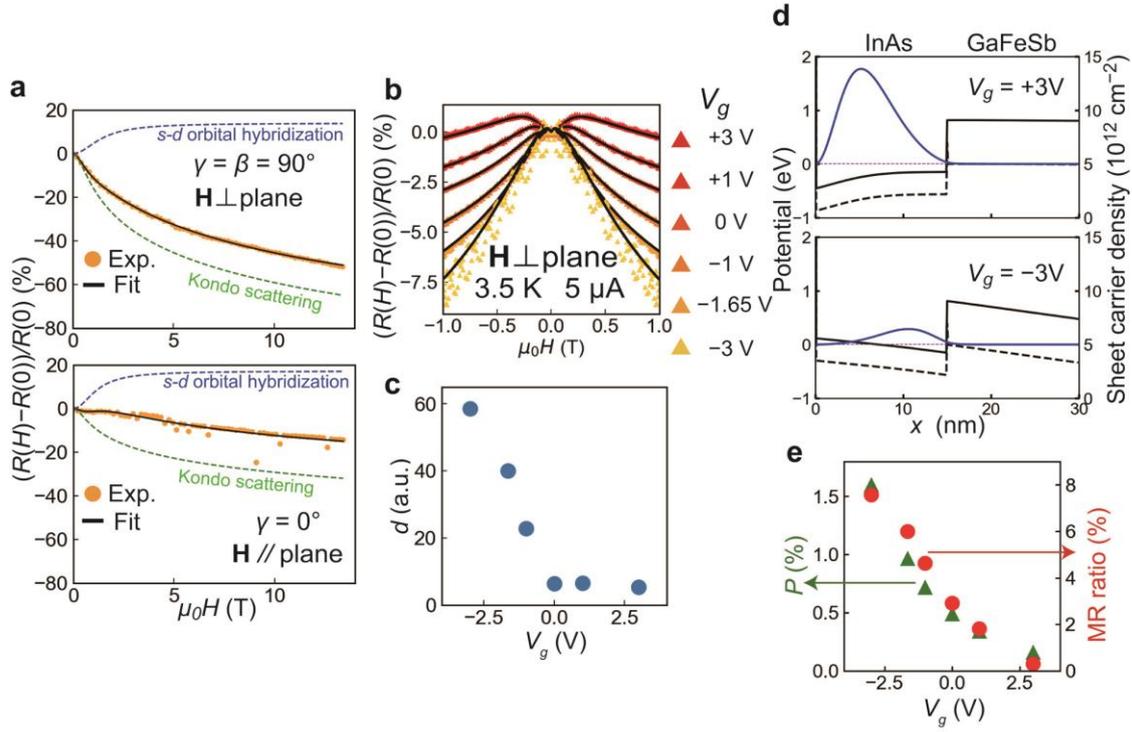

**Fig. 3| Theoretical model and gate voltage dependence of PMR. a**, The PMR [orange dots, defined as [$R(H) – R(0 \text{ T})$]/$R(0 \text{ T})$] versus magnetic field strength $\mu_0 H$ with **H** applied perpendicular (upper graph) and parallel (bottom graph) to the film plane. The black solid curves are the fitting curves obtained by eq. (1). The green and blue dashed curves are the first (Kondo scattering) and second (s-d orbital hybridization) terms in eq. (1), respectively. **b**, Evolution of the PMR ratio by varying $V_g$ from −3 V to 3 V, with $\mu_0$**H** applied perpendicular to the film plane. Black curves are the fitting curves obtained using eq. (1). **c**, $V_g$ dependence of the parameter $d$ estimated by fitting eq. (1) to the PMR data. **d**, Results obtained from the self-consistent calculation of the potential profile and the electron distribution in the InAs/(Ga,Fe)Sb bilayer of sample B at $V_g$ = 3 V (top panel) and −3 V (bottom panel), respectively (see Methods). The black solid curves, the black dashed curves, the blue curves and the purple dotted lines denote the conduction band bottom, the valence band top, the electron distribution, and quasi Fermi level $E_F$, respectively. **e**, $V_g$ dependences of $P$ (green triangles, left axis), which quantifies the penetration of the electron wavefunction into (Ga,Fe)Sb, and the PMR ratio at 1 T (red circles, right axis).

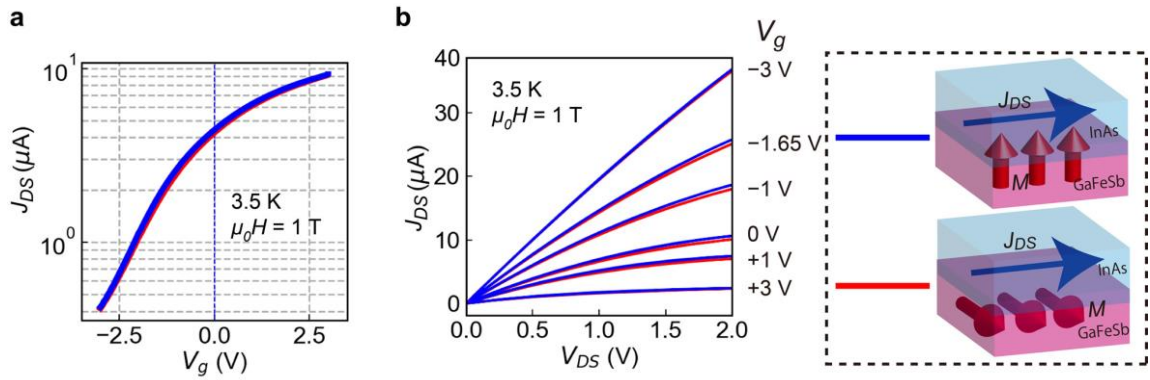

**Fig. 4| Transistor operation of the FET device fabricated on the InAs/(Ga,Fe)Sb bilayer in sample B. a**, $J_{DS} - V_g$ and **b**, $J_{DS} - V_{DS}$ characteristics at 3.5 K when the magnetization of (Ga,Fe)Sb is oriented perpendicular (blue line) and parallel (red line) to the film plane by applying a magnetic field $\mu_0 H = 1$ T.

# Supplementary Information

## Giant gate-controlled proximity magnetoresistance in semiconductor-based ferromagnetic / nonmagnetic bilayers

K. Takiguchi[1], L. D. Anh[1,2], T. Chiba[3], T. Koyama[4], D. Chiba[4], and M. Tanaka[1,5]

[1] *Department of Electrical Engineering and Information Systems, The University of Tokyo, Bunkyo-ku, Tokyo 113-8656, Japan.*
[2] *Institute of Engineering Innovation, The University of Tokyo, Bunkyo-ku, Tokyo 113-8656, Japan.*
[3] *National Institute of Technology, Fukushima College, Iwaki, Fukushima, 970-8034, Japan*
[4] *Department of Applied Physics, The University of Tokyo, Bunkyo-ku, Tokyo 113-8656, Japan*
[5] *Center for Spintronics Research Network, The University of Tokyo, Bunkyo-ku, Tokyo 113-8656, Japan.*

**Supplementary Figures**

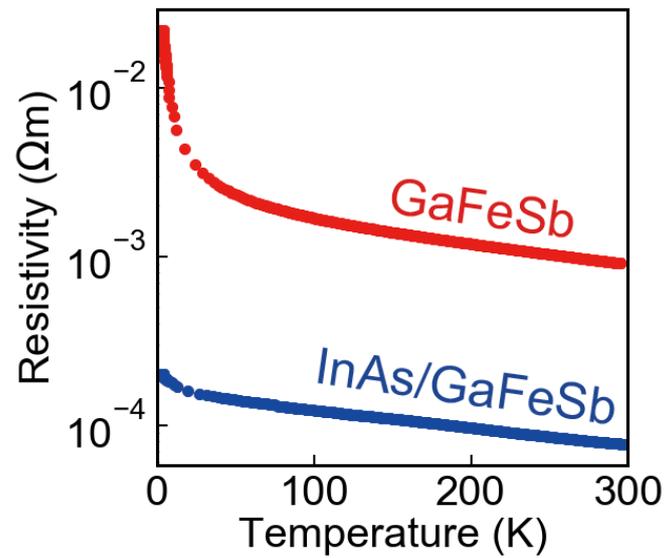

**Supplementary Figure S1.** Temperature dependence of the resistivity of an InAs (15 nm)/(Ga$_{0.8}$,Fe$_{0.2}$)Sb (15 nm) bilayer (blue circles) and a 15 nm-thick (Ga$_{0.8}$,Fe$_{0.2}$)Sb thin film (red circles). The resistivity of the InAs/(Ga,Fe)Sb bilayer is one to two orders of magnitude smaller than that of the (Ga,Fe)Sb thin film. These results imply that more than 99% of the current flows in the InAs layer, particularly at low temperature.

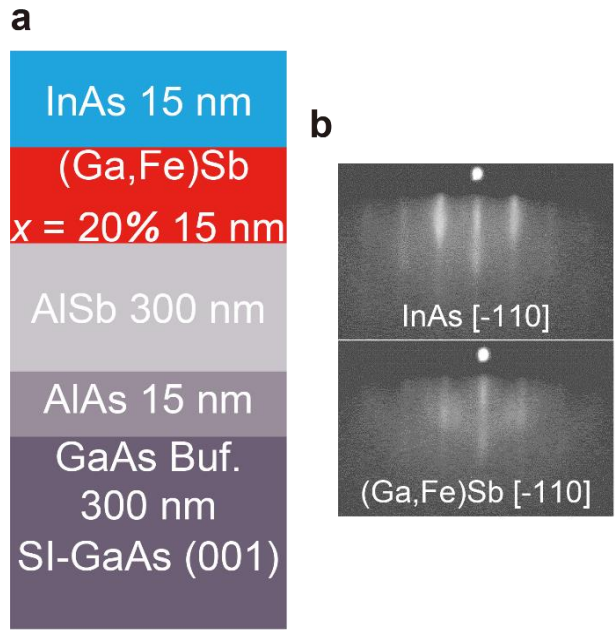

**Supplementary Figure S2. a,** Structure of the sample A and sample B presented in the main manuscript, which includes an InAs (15 nm)/(Ga,Fe)Sb bilayer. **b,** *In situ* reflection high energy electron diffraction (RHEED) patterns along the [$\bar{1}$10] axis during the MBE growth of the InAs and (Ga,Fe)Sb layers in sample A. Bright and streaky RHEED patterns indicate good crystal quality in both layers.

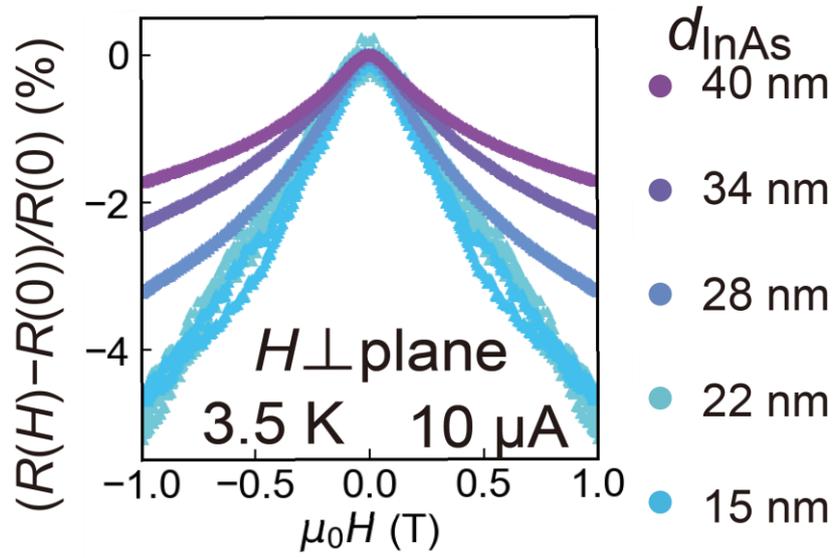

**Supplementary Figure S3.** InAs thickness dependence of the proximity magnetoresistance (PMR) measured in a series of samples consisting of InAs ($d_{InAs}$ = 15 – 40 nm)/(Ga,Fe)Sb (15 nm) bilayers at 3.5 K with $J_{DS}$ = 10 μA. The PMR ratio decreases when increasing the InAs channel thickness, but remains finite (~1.6% at 1 T) at $d_{InAs}$ = 40 nm.

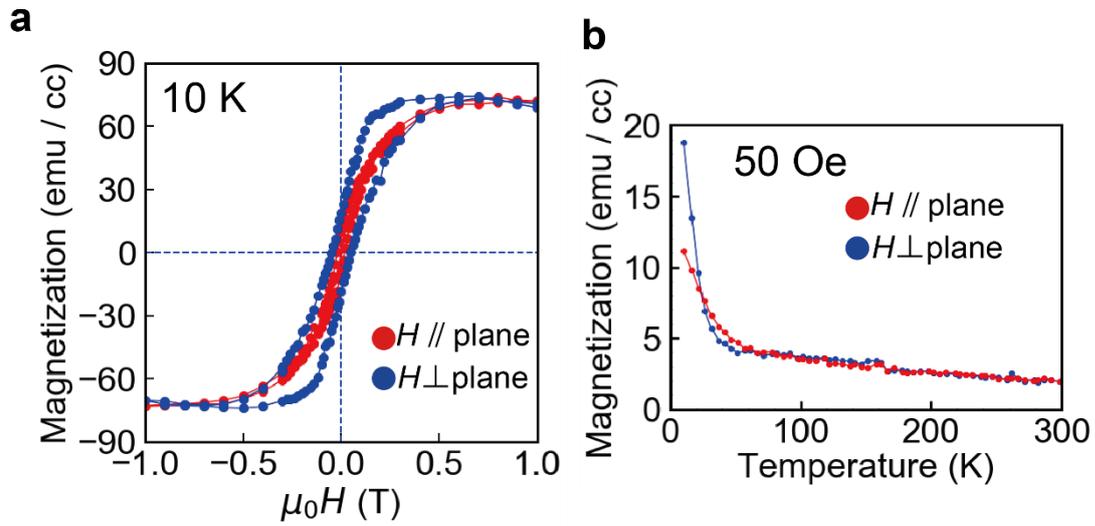

**Supplementary Figure S4.** **a,** Magnetization hysteresis curves at 10 K and **b,** magnetization (at an external field of 50 Oe) versus temperature measured in sample A measured by a superconducting quantum interference device (SQUID). In both graphs, blue and red curves correspond to the cases that the magnetic field $\mu_0\mathbf{H}$ is applied perpendicular ([001] axis) and parallel ([$\bar{1}10$] axis) to the plane, respectively.

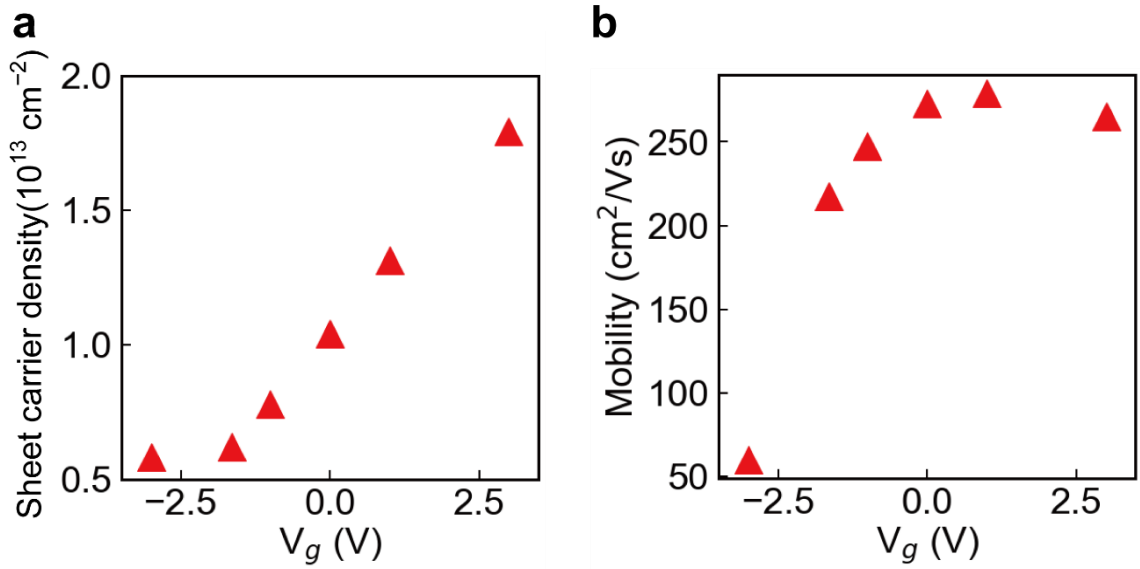

**Supplementary Figure S5.** **a,** Sheet carrier (electron) density and **b,** electron mobility of the InAs/(Ga,Fe)Sb bilayer in the FET structure fabricated on sample B as functions of the gate voltage $V_g$ at 3.5 K with $J_{DS}$ = 10 μA. When negative $V_g$ is applied, the carrier density and mobility are suppressed as the electron wavefunction is pushed towards the insulating (Ga,Fe)Sb side.

**Supplementary Note 1: Comparison between the resistivities of InAs/(Ga,Fe)Sb bilayer and (Ga,Fe)Sb single layer**

We compare the resistivity of an InAs (15 nm)/(Ga$_{0.8}$,Fe$_{0.2}$)Sb (15 nm) bilayer and a 15 nm-thick (Ga$_{0.8}$,Fe$_{0.2}$)Sb thin film in a wide range of temperature from 3.5 K to 300 K (Supplementary Figure S1). At 3.5 K, the resistivity of the (Ga,Fe)Sb thin film is two orders of magnitude larger than that of the InAs/(Ga,Fe)Sb bilayer. This indicates that more than 99% of the current flows in the InAs channel. This result is consistent with the fact that Hall measurements at all temperatures always indicate n-type conduction and the observed electron mobility is much higher than the hole mobility of (Ga,Fe)Sb. Therefore, the magnetoresistance presented in the main manuscript is a phenomenon occurring in the nonmagnetic InAs channel.

**Supplementary Note 2: InAs channel thickness ($d_{InAs}$) dependence of the proximity magnetoresistance (PMR)**

We study the dependence of the PMR on the thickness of the InAs channel ($d_{InAs}$) in a series of samples consisting of InAs ($d_{InAs}$ = 15 - 40 nm)/(Ga,Fe)Sb (15 nm) bilayers, as shown in Supplementary Figure S3. The PMR decreases when increasing $d_{InAs}$. This behavior supports our conclusion that the PMR observed in the InAs/(Ga,Fe)Sb bilayers is originated from the magnetic proximity effect (MPE) at the bilayer interface, whose effect becomes weaker at thicker $d_{InAs}$. The PMR remains finite (~1.6% at 1 T) even at $d_{InAs}$ = 40 nm, which indicates long-range magnetic coupling in the InAs/(Ga,Fe)Sb bilayers.

**Supplementary Note 3: Magnetic properties of the InAs/(Ga,Fe)Sb bilayers**

We characterized the magnetic properties of the InAs/(Ga,Fe)Sb bilayer in the sample A by SQUID. The measurements were carried out after cooling the sample while applying an external magnetic field $\mu_0 H$ (= 1 T). The magnetization of the (Ga,Fe)Sb layer in sample A measured with $\mu_0 H$ applied parallel (along the GaAs [$\bar{1}$10]) and perpendicular (along the GaAs [001]) to the film plane, respectively, shows clear hysteresis loops at 10 K (Supplementary Figure S4a). For measuring the temperature dependence of magnetization (Supplementary Figure S4b), a small magnetic field of 50 Oe was applied. In both field directions, the magnetization remains finite at above 300 K. This means that the Curie temperature $T_C$ of the (Ga,Fe)Sb layer in sample A is higher than room temperature.

**Supplementary Note 4: Sheet carrier density and mobility in the InAs/(Ga,Fe)Sb bilayer as functions of the gate voltage**

Supplementary Figure S5 shows the sheet carrier (electron) density (a) and electron mobility (b) of the InAs/(Ga,Fe)Sb bilayer in the FET structure fabricated on sample B as functions of the gate voltage $V_g$ at 3.5 K. The sheet carrier density was estimated from the Hall measurements at various $V_g$ values, which always indicate n-type conduction. When negative $V_g$ is applied, the carrier density and mobility are decreased. As we mentioned in the main text, the calculation of the band profile and the electron distribution shows that the negative $V_g$ leads to large penetration of the electron wavefunction in the InAs QW into the ferromagnetic layer (Ga,Fe)Sb. The large penetration of the carrier wavefunction into the (Ga,Fe)Sb layer causes stronger interfacial scattering, which is consistent with the smaller mobility at negative $V_g$.